\documentstyle{article}

\title{{\bf Cluster, Backbone and Elastic Backbone Structures of 
the Multiple Invasion Percolation}} 

\author{Roberto N. Onody and Reginaldo A. Zara\\ \\ 
\small {\em Departamento de F\'{\i}sica e Inform\'{a}tica } \\
\small {\em Instituto de F\'{\i}sica de S\~{a}o Carlos} \\
\small {\em Universidade de S\~{a}o Paulo - Caixa Postal 369} \\
\small {\em 13560-970 - S\~{a}o Carlos, S\~{a}o Paulo, Brasil.}}
\date{}
\begin{document}
\maketitle
\normalsize
\baselineskip=24pt
\begin{abstract}
We study the cluster, the backbone and the elastic backbone structures
of the multiple invasion percolation for both the perimeter and
the optimized versions. We investigate the behavior of the
mass, the number of red sites (i. e., sites through which all
the current passes) and loops of those
structures. Their corresponding scaling exponents are also estimated. 
By construction, 
the mass of the optimized model scales exactly with the gyration
radius of the cluster - we verify that this also happens to the
backbone. Our simulation shows that the red sites almost disappear,
indicating that the cluster has achieved a high degree of
connectivity.
\\ \\ \\ \\ \\

\end{abstract}

\newpage

\section{INTRODUCTION}

When a nonviscous liquid is injected into a porous medium already
filled with a viscous fluid two distinct regimes appear: one where
the dominant forces are of {\em capillary} nature and another
where the {\em viscous} forces are predominant. Depending on the
injection rate the system can be found in one of these regimes. The
theoretical description of such a system is based on two models:
the invasion percolation \cite{wil-wil} and the diffusion-limited
aggregation (DLA) \cite{wit-san}. The invasion percolation model
is indicated when the fluid flow is slow, that is, when the capillary
number is small. The displacement process of the fluid follows
minimum resistance paths: the smaller pores are filled or invaded
first. 

Grassberger and Manna \cite{gra} pointed out that the invasion
percolation is a kind of self-organizing criticality \cite{bak}
exhibiting scale invariant behavior in time and space and evolving
into a natural critical state. Indeed, there are two kinds of
invasion percolation models: {\em with} and {\em without} trapping 
\cite{wil-wil}. The trapping occurs when the displaced fluid is an
uncompressible fluid and it is completely surrounded by the other.
These models belong to different universality class. The version
with trapping has a fractal dimension $DF \sim 1.82$ and the case
without trapping corresponds to the {\em critical ordinary
percolation} \cite{wil-wil} ($DF = \frac{91}{48}$). Important
applications of the invasion percolation model were found, extending
from the terciary recovery of petroleum to the fingering phenomena
in soils \cite{ono1}.

Many modifications of the original invasion percolation model have
been proposed. They take into account the action of an external
gravitational field \cite{wil,bir,mea} or the flux with a privileged
direction \cite{ono2}. In the pioneer formulation of the invasion
percolation model \cite{wil-wil}, at each growth step only
{\em one} lattice site was allowed to be occupied. Recently \cite
{ono3}, a more realistic model was investigated which permits that
a certain number of lattice sites can be invaded at the same time:
the {\em multiple invasion percolation} model. There are two kinds
of multiple invasion percolation: the perimeter model and the
optimized model. In the first model the cluster growth is controlled
by the flux through the perimeter. The optimized model is governed
by a scaling relation between the mass and the gyration
radius of the cluster. Reference \cite{ono3} studied the multiple
invasion percolation (in its {\em site} version, as in this paper)
determining the abundance of vertice type, the mean coordination
number, the acceptance profile and the fractal dimensions. An
interesting {\em burst} phenomenon was detected and analyzed in the
optimized model.  

The backbone is the intersection of {\em all} self-avoiding walks
connecting two points $P_{1}$ and $P_{2}$ of the lattice. This
means that if we pass a current between $P_{1}$ and $P_{2}$ the
backbone is the set of points carrying current, all dangling ends
are discarded. The elastic backbone is the union of all the
shortest paths between $P_{1}$ and $P_{2}$. In our case $P_{1}$ is
the lattice center and $P_{2}$ is the point where the cluster find
the frontier for the first time ( the growth process stops at
this moment). The investigation of the backbone of clusters has
been of interest for a long time. Possible applications are the
conductivity of random systems \cite{gen} and the flow of fluids
in porous media \cite{sta-co}.

The cluster, the backbone and the elastic backbone are the important
{\em structures} of the fractal objects. The determination of
the properties of such structures can lead to a better understanding
of the fractal objects and even to a classification scheme
for them. But what are the relevant parameters to be measured in
these structures?. We can list the following quantities: the mass,
the minimum path, the number of red points and the number of loops.
At criticality, all them scale as a power law with the lattice
size. So they can be charactherized by their corresponding scaling
exponents. 

The minimum path is the shortest distance between two lattice
points. The lengths of the minimum path or `chemical distance' are
usually greater than their Euclidean distance \cite{he-sta}.
The red points are the throttle points through which all the
current pass - if they are removed the flow stops.

In the present paper we study the cluster, the backbone and
the elastic backbone structures of the multiple invasion percolation
for both the perimeter and optimized models. To determine
the backbone and the elastic backbone we employed the burning
algorithm \cite{herr}. Although there are actually more efficient
algorithms based on artificial intelligence theory \cite{don} or
recursive algorithm \cite{gras}, we prefer the burning technique
because beyond the backbone and elastic backbone identification
it also permits the determination of the red sites and loops.

The scaling exponents for the mass, the red sites, the minimum
path and the loops are found for many values of the parameters $F$
and $D$ of the perimeter and the optimized models. We did not
strive to make these exponents very precise. Indeed we payed more
attention on the physical changes coming up with the variation of
the parameters, as long as both models were conceived to
continously interpolate from fractal to compact objects. 
The optimized model reveals
two amazing properties: not only the cluster but also the backbone
mass scales {\em exactly} with the gyration radius and the red
points practicaly do not exist anymore. This means that the cluster,
generated with the optimized algorithm, has acquired a high degree
of connectivity without having to increase its fractal dimension.

\section{THE PERIMETER MODEL}

We briefly recall the growth mechanism established for the perimeter
model. Suppose that at some growth stage $t$ the cluster mass is
$M_{t}$ and that the rectangle area inside which the cluster is
inscribed is $A$. The square root of $A$ can be interpreted as a
measure of the correlation length \cite{ono4}. This interpretation
comes from the fact that, as in the ordinary invasion percolation,
the multiple invasion percolation can also be thought as a kind of
critical percolation model \cite{ono3}. At time $t+1$, the cluster
mass $M_{t+1}$ will be given by

\begin{equation}
M_{t+1} = M_{t} + INT(4 F \sqrt{A})
\label{s2e1}
\end{equation}
where INT means the integer part, $F$ is an external parameter
$( 0 \leq F \leq 1)$ corresponding to the fraction of the perimeter
$4 \sqrt{A}$ to be invaded at time $t+1$. We start the growing
process at the center of a square lattice.

It was numerically shown in reference \cite{ono3} that for $F$
values greater than 1/2 the cluster is compact and for
$(0 \leq F \leq 0.5)$ it interpolates between the ordinary
invasion percolation (with fractal dimension $ DF =\frac{91}{48}$)
and the closed packed limit ($DF=2$). We found a simple analytic
demonstration of this fact. For a lattice size $L$, a cluster
growing {\em compactly} from its center will touch the boundary at
a time $t=\frac{L}{2}$ and it will acquire the form of a square
with side $l=\frac{L}{\sqrt{2}}$. The mass is $M_{t}=\frac{L^{2}}{2}$
. At time $t+1$, {\em all} available sites will be invaded and
the maximum possible mass is $M_{t+1}=\frac{(L+2)^2}{2}$.Using
(\ref{s2e1}) we have the inequality

\begin{equation}
F \leq \frac{1}{2} + \frac{1}{2L}
\label{s2e2}
\end{equation}
which for large lattices saturate at $F=\frac{1}{2}$. 

In order to obtain the scaling exponents, throughout this section
we use lattices size $L=$ 51, 101, 201 and 401.

\subsection{The Mass}

The mass of fractal objects  \cite{man} scales with the lattice
size $L$ as
  
\begin{equation}
M \sim L^{DF}
\label{s2e3}
\end{equation}

The cluster mass, the fractal dimension and its dependence on $F$
were already studied \cite{ono3}. Here we extend the results to the
backbone and the elastic backbone. The data for the backbone are of
good quality and they were obtained averaging over 100-2000
realizations. We get, for example, $DF(F=0) = 1.64 \pm 0.01$ which is
completely compatible with the most extensive simulation performed
by Grassberger \cite{gras} who got $ 1.647 \pm 0.004 $. Our results
are shown in Table 1. With increasing $F$ the backbone
fractal dimension goes to 2 in a faster way than those of the cluster
itself. From Table 1 we see that around $F \sim 0.3$ some exponents
break their monotone behaviour. At this point the cluster has a
circular form and the corresponding gyration radius is maximum
\cite{ono3}.

For the elastic backbone we found deviations from the straight line
when we plotted ln($M$) x ln($L$). This strongly indicates that
corrections to scaling are necessary. We adopted the ratio method
\cite{he-sta} to correct them. For $F=0$ we got $DF = 1.17 \pm 0.04$
which is in fair agreement with Herrmann {\em et al.} \cite{herr}
result: $ 1.10 \pm 0.05$. When $F$ increases, the trend is that the
elastic backbone approaches the form of a straight line connecting
the lattice center to the point where the cluster hits the frontier.
The mass exponent goes to one.

\subsection{The Red Sites}

The number of red sites $N_{r}$ scales as

\begin{equation}
N_{r} \sim L^{D_{r}}
\label{s2e4}
\end{equation}

Here again corrections to scaling are necessary. Of course, the
exponents $D_{r}$ for the cluster and the backbone are the same.
In the case of the ordinary invasion percolation ($F=0$), $D_{r}$
is known exactly \cite{aco}. Coniglio used the relations between
the percolation model, Potts model and Coulomb gas to get
$D_{r}= \frac{3}{4}$. Our result $D_{r}(F=0)=0.77 \pm 0.02$ is in good
agreement. As $F$ goes to $0.5$ the exponent $D_{r}$ approaches
zero (see Table 1). For the elastic backbone we get $D_{r}(F=0)=
1.08 \pm 0.05$. As we have already observed, the elastic backbone
approaches a straight line with increasing $F$. This means that
almost every site belonging to the elastic backbone is a red site,
so $D_{r} \sim DF \rightarrow 1$ as $F$ goes to $0.5$.

\begin{table}
\begin{center}
\caption{\label{s2t1} }
{\em To be inserted.}
\end{center}
\end{table}
 
\subsection{The Loops}

We can put a bond connecting any two nearest neighbors occupied sites.
The result is a connected graph for which the Euler relation
holds: $ N_{l}= N_{b}-M+1$ , where $ N_{l} $ is the number of cycles or
loops number; $ N_{b} $ is the number of bonds and $ M $ is the mass or
the number of sites. For the burning algorithm on the square
lattice, $ N_{l} $ is calculated by counting the number of times that
one tries to burn a site that is already burned in the same time unit.

The number of loops $N_{l}$ scales with the lattice size $L$ as

\begin{equation}
N_{l} \sim L^{D_{l}}
\label{s2e5}
\end{equation}

The data are of very good quality and no correction to scaling was
necessary. The exponent $D_{l}$ approaches $2$ with increasing $F$
for both the cluster and the backbone. This is a consequence of
the Euler relation. 
As $ F $ increases, the clusters become more compact, $ N_{b} $
approaches $ 2M $ and, for large clusters, $N_{l} \sim M \sim L^{2}$.
On the other hand, looking at Table 1, the
exponent seems to approach $1$ for the elastic backbone but this
not true. In Fig.1 we show one typical cluster with $F=0.4$ and
$L=401$. The elastic backbone starts at the center and follows to
the right nearly as a straight line until it finds an obstacle
(a site with a large random number associated). Then a kind of a
jet appears which increases the number of loops. In the limit
$ L \rightarrow \infty $ such effect can only be avoided if
$F \geq 0.5$. This means that for the elastic backbone $D_{l}$ goes
abruptly to zero near $F \sim 0.5$ as it is shown in the inset
of Fig.1.

\begin{figure}
\begin{center}
\caption{\label{s2f1} }
{\em To be inserted.}
\end{center}
\end{figure}

\subsection{The Minimum Path}

The minimum path scales as 

\begin{equation}
l_{min} \sim L^{D_{min}}
\label{s2e6}
\end{equation}

Naturally, the exponent $D_{min}$ is the same for both the cluster
, backbone and elastic backbone. Our estimate
$D_{min}(F=0) = 1.14 \pm 0.04$ is consistent with the most precise
value $D_{min}= 1.1307 \pm 0.0004$ \cite{gras}. We see from Table 1
that $D_{min}$ approaches one with increasing $F$.

\section{THE OPTIMIZED MODEL}

The optimized model \cite{ono3} was devised in order to have a
growth mechanism obeying {\em exactly} the scaling
 
\begin{equation}
M \sim (Rg)^{D} 
\label{s3e1}
\end{equation}
or as near it as possible ($Rg$ is the gyration radius and $D$ is
a real positive external parameter that can be tuned). 

Basically we use 
the following strategy: at {\em each} growing step we have a list
containing {\em all} the cluster perimeter sites that can be invaded
and we ask for the number of sites that should be invaded in order
that (\ref{s3e1}) is verifyed as closely as possible . This proceeding
builds a fractal object which is extremely stabilized in the
sense that in {\em any} stage or size the scaling is perfectly
obeyed not only in the asymptotic limit (as usually). Another
important advantage is that the necessity of mass averages on the 
cluster ensemble diminishes. We hope this can be very
useful in dilute systems. 

Any cluster is very representative because
the mass dispersion is very small. As an example, we compare the 
ratio $\frac{N_{p}}{N_{o}}$ ($N_{p}$ and $N_{o}$ are the number of 
experiments
performed in the perimeter and optimized models, respectively) when
both models are simulated to achieve nearly the same relative
standard deviation $\frac{\Delta M}{M}$. The results are shown in
Table 2. 

\begin{table}
\begin{center}
\caption{\label{s3t1} }
{\em To be inserted.}
\end{center}
\end{table} 

For one to one
realization,   we  have  already   compared  the   optimized
algorithm to that of the ordinary invasion percolation.  The
results were even more  impressive  ( see Fig.6 of ref \cite{ono3}).

As shown in reference \cite{ono3}, when $D \in [\frac{91}{48},2]$
it coincides with the usual fractal dimension $DF$; if
$ 0 < D < \frac{91}{48}$ the system is frustrated in the sense that
it tries but fails to invade less than one site; if $D > 2$ the system
is also frustrated but now for a different reason: the invasion cannot
be faster than it is allowed by the two dimensional lattice.
In the last two situations $D \neq DF$.

In this paper we use the optimized algorithm to simulate only {\em
one} cluster. The growth of this unique cluster is stopped at each
$50$ steps. Then we measure the mass and the loops number for both
the cluster and its backbone. Their corresponding gyration radius
and the minimum path are also determined. At each stage the rectangle
in which the cluster is actually inscribed is also obtained (up
to lattice size $L=401$). The lattice center and the point $P_{2}$
where the cluster touches that rectangle are used to get the minimum
path. The elastic backbone is the collection of all these paths.
It is a faint structure with almost one dimensional characteristics.
When we pass from one rectangle to the next, usually happens that
$P_{2}$ turns, for example, from north to south or east. So in just
one step the sites composing the elastic backbone change wildly,
precluding its determination (remember that here no averages are
made).  

We studied the optimized model only in the physical region
$D \in [\frac{91}{48},2]$. Below we present our results.

\subsection{The Mass}

Naturally, $D \equiv DF$ for the cluster mass. To our surprise the
scaling (\ref{s3e1}) is also perfectly obeyed by the backbone (see
Fig.2(a)). Remember that the equation (\ref{s3e1}) was only imposed
on the cluster. In Table 3 we present the fractal dimensions $DF$
for some values of $D$. Remember that we performed only a unique
realization. The errors bars correspond to the standard deviation
calculated along this experiment. Observe that at $D=1.89$ (the fractal
dimension of the ordinary invasion percolation) we get for the
backbone $DF=1.74 \pm 0.01$ which is greater than the $ 1.647 \pm
0.004 $ \cite{gras} of the ordinary invasion percolation. This
means that although the optimized model at $D=1.89$ and the
ordinary invasion have the {\em same} cluster fractal dimension,
they are intrinsicaly different since their backbones are different.
Conductivity properties will not be the same.

\begin{table}
\begin{center}
\caption{\label{s3t2} }
{\em To be inserted.}
\end{center}
\end{table}

\subsection{The Red Sites}

An astonishing result that we got is that the number of red sites
$N_{r}$ of the optimized model is {\em very small and random}.
It does not obey any power law. For example, when we grow one
cluster and count the red sites number at sizes $L=51, 101, 151,
201, 251, 301, 401$ we find $N_{r}= 1, 8, 8, 1, 6, 3, 7$. 
To investigate this further, we simulate at $D=1.95$
(just in the middle of the physical region $[1.89,2.00]$) with 
lattice sizes $L=51,101,151,201$ and the number of realizations 
$100,100,60,15,$ respectively. We got $<N_{r}>=1.33,1.49,1.43,1.87$.
This bring us to the conclusion that for our optimized model
the concept of red sites is not important. The red sites number 
is so small that the probability of disconnecting the cluster by
removing randomly any site is pratically zero. The optimized 
algorithm destroys the red sites increasing the cluster connectivity.

\subsection{The Loops}

The number of loops $N_{l}$ scales with the gyration radius $Rg$
in the usual way 

\begin{equation}
N_{l} \sim (Rg)^{D_{l}}
\label{s3e2}
\end{equation}

Looking the Table 3 we conclude that, as expected, both exponents 
go to 2 with increasing $D$. 

\begin{figure}
\begin{center}
\caption{\label{s3f1} }
{\em To be inserted.}
\end{center}
\end{figure}

To see the influence of the ensemble averages on the scaling
exponents we have  performed $40$
realizations of the optimized  model on a lattice size $L=251$
and $D=1.91$. We got $D_{l}=1.73(4)$ for the cluster;
$DF=1.84(1)$ and $D_{l}=1.70(2)$ for the backbone and $D_{min}=
1.02(3)$. These results are fairly good when compared to 
$D_{l}=1.78(1)$; $DF=1.82(1)$ and $D_{l}=1.72(1)$ and $
D_{min}=1.05(2)$, respectively,
obtained for just one realization on size $L=401$. Unfortunately,
we were not able to simulate larger lattices face to the huge
CPU demand ( a unique $L=401$ realization took 52 hours on a
Alpha/275 station).

\subsection{The Minimum Path}

The minimum path scales as (\ref{s2e6}). Our results are shown in
Table 3. Like in the perimeter model, the exponent $D_{min}$
approaches 1 as the cluster becomes more compact. 

\subsection{The Burst Phenomenon}

We already know from reference \cite{ono3} that
the system is frustrated when $ D \in [0,\frac{91}{48}]$ ($DF=
\frac{91}{48}$) or $D > 2$ ($DF=2$). In the last regime the burst
phenomenon takes place. This corresponds to an enormous and sudden 
mass explosion. We simulate at $D=5.00$ ($L=201$) just be
fore and after one of such explosions ( at time step 965,
just as in Fig.11 of the reference \cite{ono3} ). The results for
the cluster and the backbone are shown in Table 4. It shows a
dramatic increase (decrease) of the mass and loops (red sites).

\begin{table}
\begin{center}
\caption{\label{s3t3} }
{\em To be inserted.}
\end{center}
\end{table}

\section{CONCLUSIONS}

We use the burning method to identify and analyze the cluster, the
backbone and the elastic backbone structures of the multiple
invasion percolation model. We determine the scaling exponents
for both the perimeter and the optimized models as well as 
their dependence with the parameters $F$ and $D$. For those
structures we also study the behaviour of the mass, the number of red
points, the number of loops and the minimum path.
The optimized model in the physical
region  $ D=DF \in [\frac{91}{48},2]$ exhibited two amazing
properties: the perfect scaling of the backbone mass with its gyration
radius and the red points disappearance. This model seems to be well 
suited to treat dilute systems where the fluctuations of the clusters 
ensemble hamper the data accuracy and overcast the reality.

\vspace{1cm}
\noindent {\bf Acknowledgments }\\ \\
We acknowledge CNPq (Conselho Nacional de Desenvolvimento
Cient\'{\i}fico e Tecnol\'ogico) and FAPESP ( Funda\c c\~ao de Amparo
a Pesquisa do Estado de S\~ao Paulo ) for the financial support.

\newpage

\begin{center}

{\bf Tables Caption}

\end{center}

\vspace{1.5cm}

{\bf Table 1} - The scaling exponents of the perimeter model.
The order they appear in this table corresponds to the cluster,
backbone and elastic backbone respectively. Those marked with an
asterisk were calculated using the ratio method.

\vspace{1cm}

{\bf Table 2} - The ratio $\frac{N_{p}}{N_{o}}$ measures how
many times the number of experiments in the perimeter model $N_{p}$
exceeds that of the optimized $N_{o}$ when we impose they have the same
relative standard deviation $\frac{\Delta M}{M}$. The first and
the second lines correspond to the cluster and the backbone,
respectively. We used $ N_{o}(L=51,101)=100, N_{o}(L=201)=15 $.
 
\vspace{1cm}

{\bf Table 3} - The scaling exponents of the optimized model.
The order they appear in this table corresponds to the cluster
and backbone respectively. 

\vspace{1cm}

{\bf Table 4} - The abrupt variation of the mass, loops and red
points at the moment of the explosion.

\newpage

\begin{center}

{\bf Figures Caption}

\end{center}

\vspace{1.5cm}

{\bf Fig.1} - In gray is a typical cluster of the perimeter model
with $F=0.4$ on a lattice of size $L=401$. The elastic backbone
is shown in black. The inset gives the dependence of the loops
number scaling exponent with $F$.

\vspace{1cm}

{\bf Fig.2} - (a) The symbol $o$ stands for the logarithm plot
of the backbone mass versus its gyration radius. (b) The logarithm
dependence of the backbone loops number $N_{l}$ with the gyration
radius.

\newpage

\begin{table}
\begin{center}
\begin{tabular}{||c|c|c|c|c||}
\hline\hline
\multicolumn{1}{||c|}{\bf $F$} &
\multicolumn{1}{c|}{\bf $DF$} &
\multicolumn{1}{c|}{\bf $D_{r}$} &
\multicolumn{1}{c|}{\bf $D_{l}$} &
\multicolumn{1}{c||}{\bf $D_{min}$} \\
\hline 
$ $ & $1.88(1)$ & $0.77(2) ^{\ast}$ & $1.98(1)$ & $ $ \\ 
\cline{2-4}
$0.0$ & $1.64(1)$ & $0.77(2) ^{\ast}$ & $1.72(1)$ & $ 1.14(4)
^{\ast}$ \\ \cline{2-4}
$ $ & $1.17(4)^{\ast}$ & $1.08(5) ^{\ast}$ & $1.19(1)$ & $ $ \\
\hline
$ $ & $1.98(1)$ & $0.38(6) ^{\ast}$ & $1.99(1)$ & $ $ \\ 
\cline{2-4}
$0.1$ & $1.89(2)$ & $0.38(6) ^{\ast} $ & $1.92(1)$ & $ 1.05(1)
^{\ast} $ \\ \cline{2-4}
$ $ & $1.05(2)^{\ast}$ & $1.10(2) ^{\ast}$ & $1.17(1)$ & $ $ \\
\hline
$ $ & $1.99(1)$ & $0.29(6) ^{\ast}$ & $1.99(1)$ & $ $ \\
\cline{2-4}
$0.2$ & $2.00(2)$ & $0.29(6) ^{\ast} $ & $2.00(2)$ & $ 1.05(1)
^{\ast} $ \\ \cline{2-4}
$ $ & $1.02(2)^{\ast}$ & $1.08(2) ^{\ast}$ & $1.11(1)$ & $ $ \\
\hline
$ $ & $1.98(1)$ & $0.07(4) ^{\ast}$ & $1.95(1)$ & $ $ \\ 
\cline{2-4}
$0.3$ & $2.00(1)$ & $0.07(4) ^{\ast} $ & $1.94(3)$ & $1.04(1)
^{\ast} $ \\ \cline{2-4}
$ $ & $1.03(3)^{\ast}$ & $1.03(3) ^{\ast}$ & $1.12(1)$ & $ $ \\
\hline
$ $ & $1.99(1)$ & $0.05(4) ^{\ast}$ & $1.97(1)$ & $ $ \\ 
\cline{2-4}
$0.4$ & $2.00(1)$ & $0.05(4) ^{\ast} $ & $1.97(4)$ & $1.02(1)
^{\ast} $ \\ \cline{2-4}
$ $ & $1.01(4)^{\ast}$ & $1.01(3) ^{\ast}$ & $1.00(3)$ & $ $ \\
\hline\hline
\end{tabular}
\end{center}
\end{table}
\begin{center}
{\bf TABLE 1}
\end{center}

\newpage

\begin{table}
\begin{center}
\begin{tabular}{||c|c|c||}
\hline\hline
\multicolumn{1}{||c|}{\bf $L$}&
\multicolumn{1}{c|}{\bf $\frac{\Delta M}{M}$}&
\multicolumn{1}{c||}{\bf $\frac{N_{p}}{N_{o}}$} \\
\hline
$51$ & $0.009$ & $21$ \\ \cline{2-3}
$ $ & $0.013$ & $10$ \\
\hline
$101$ & $0.009$ & $24$ \\ \cline{2-3}
$ $ & $0.013$ & $13$ \\
\hline
$201$ & $0.017$ & $36$ \\ \cline{2-3}
$ $ & $0.022$ & $25$ \\
\hline\hline
\end{tabular}
\end{center}
\end{table}
\begin{center}
{\bf TABLE 2}
\end{center}

\newpage

\begin{table}
\begin{center}
\begin{tabular}{||c|c|c|c||}
\hline\hline
\multicolumn{1}{||c|}{\bf $D$}&
\multicolumn{1}{c|}{\bf $DF$}&
\multicolumn{1}{c|}{\bf $D_{l}$} &
\multicolumn{1}{c||}{\bf $D_{min}$} \\
\hline
$1.89$ & $1.89$ & $1.74(1)$ & $1.17(8)$ \\ \cline{2-3}
$ $ & $1.74(1)$ & $1.63(1)$ & $ $ \\
\hline
$1.91$ & $1.91$ & $1.78(1)$ & $1.05(2)$ \\ \cline{2-3}
$ $ & $1.82(1)$ & $1.72(1)$ & $ $ \\
\hline
$1.95$ & $1.95$ & $1.83(1)$ & $1.00(2)$ \\ \cline{2-3}
$ $ & $1.93(1)$ & $1.82(1)$ & $ $ \\
\hline
$2.00$ & $2.00(1)$ & $2.00(1)$ & $1.00(1)$ \\ \cline{2-3}
$ $ & $1.99(1)$ & $1.99(1)$ & $ $ \\
\hline\hline
\end{tabular}
\end{center}
\end{table}
\begin{center}
{\bf TABLE 3}
\end{center}

\newpage

\begin{table}
\begin{center}
\begin{tabular}{||c|c|c|c|c||}
\hline\hline
\multicolumn{1}{||c|}{}&
\multicolumn{2}{c|}{\bf Cluster}&
\multicolumn{2}{c||}{\bf Backbone} \\
\hline
$ $ & $before$ & $after$ & $before$ & $after$ \\ \hline
$mass$ & $1619$ & $2358$ & $1045$ & $2310$ \\ \hline
$red points$ & $52$ & $1$ & $52$ & $1$ \\ \hline
$loops$ & $863$ & $1981$ & $776$ & $1981$ \\ 
\hline\hline
\end{tabular}
\end{center}
\end{table}
\begin{center}
{\bf TABLE 4}
\end{center}

\newpage

\begin{center}
FIGURE 1
\end{center}
\begin{figure}[b]
\includegraphics{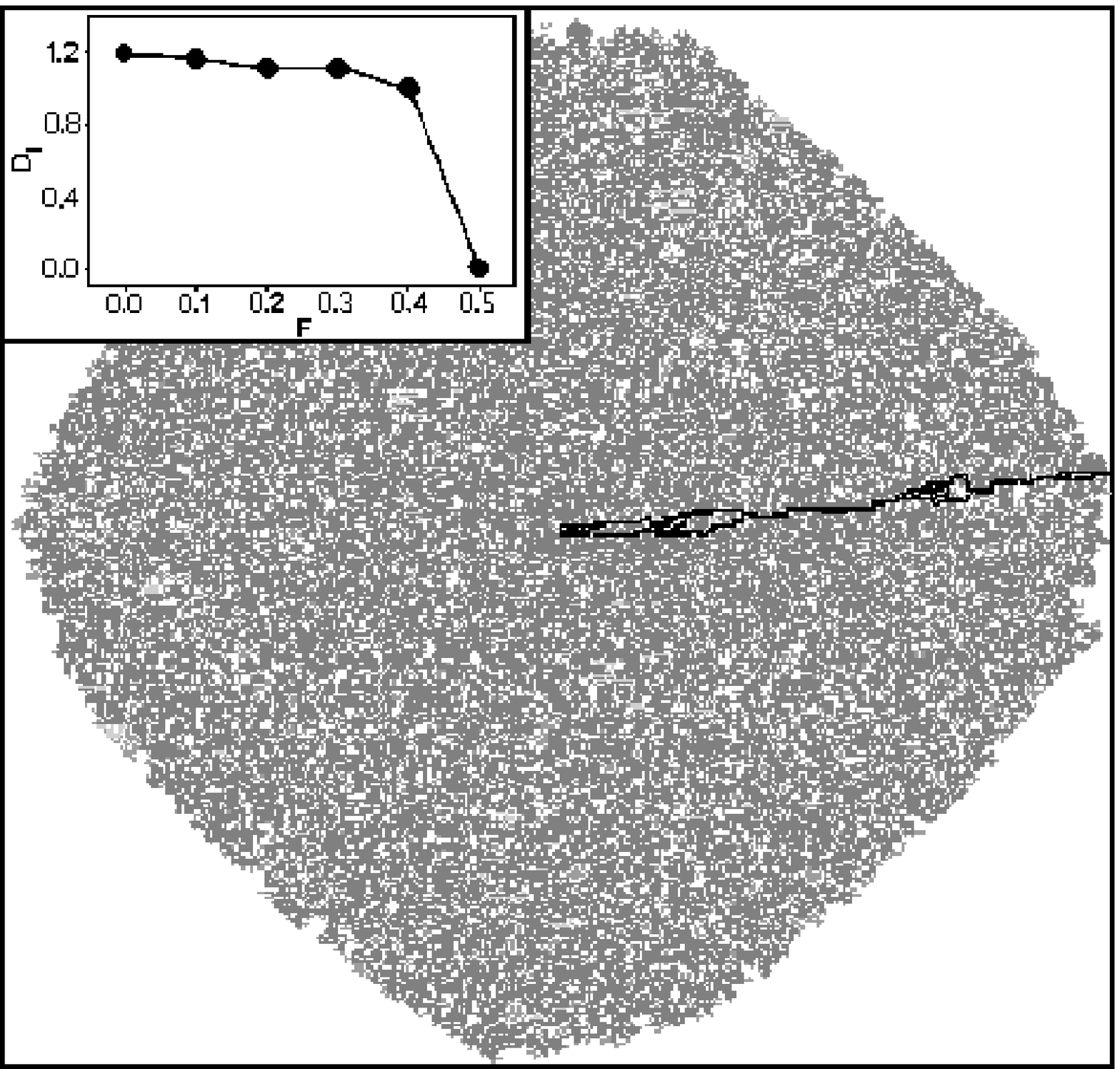}
\end{figure}

\newpage

\begin{center}
FIGURE 2
\end{center}
\begin{figure}[b]
\includegraphics{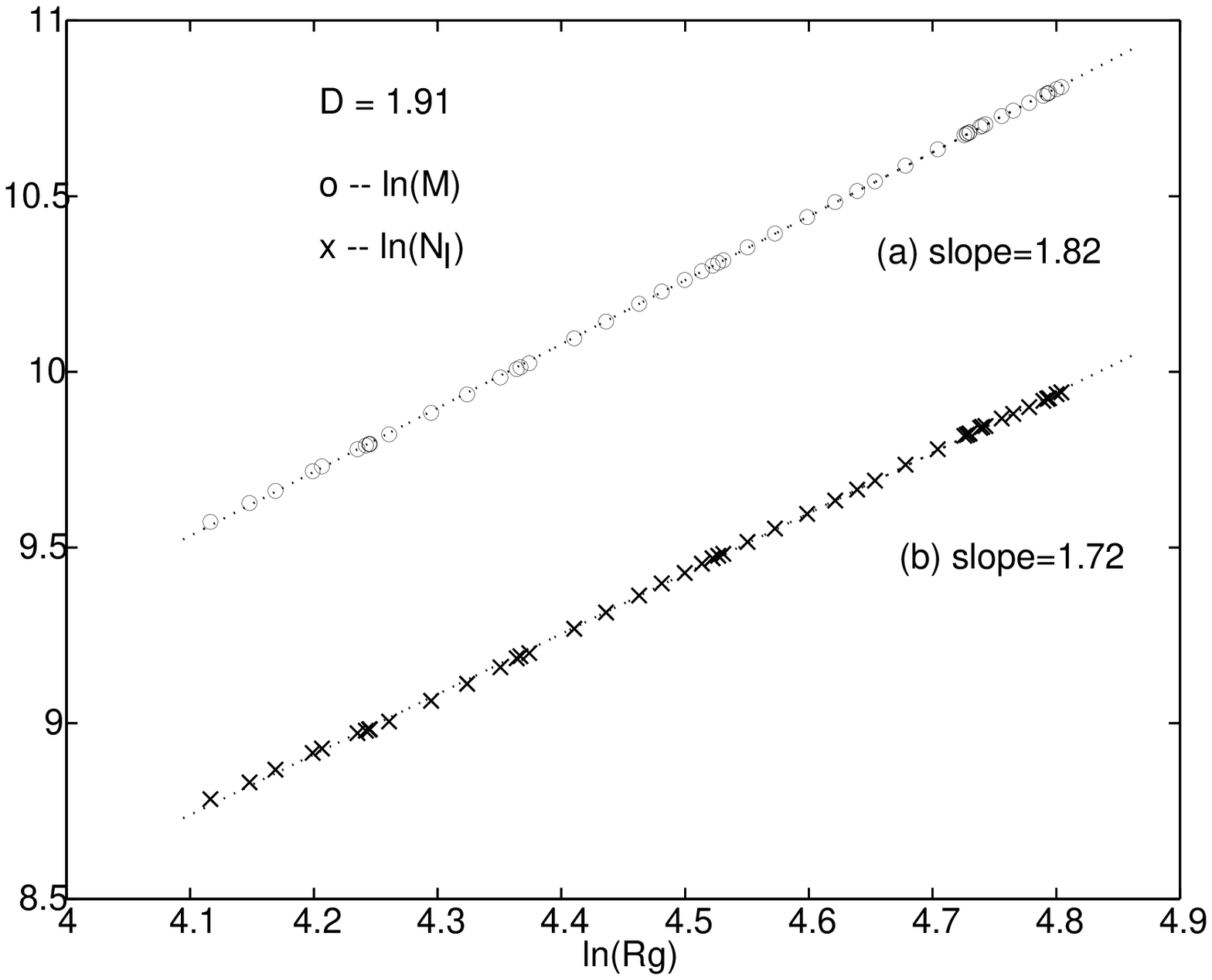}
\end{figure}

\end{document}